\newbox\leftpage
\newdimen\fullhsize
\newdimen\hstitle
\newdimen\hsbody
\tolerance=1000\hfuzz=2pt
\overfullrule=0pt
\def\bigans{b }
\def\answ{b }
\ifx\answ\bigans\message{(this will come out unreduced.}


\magnification=1200\baselineskip=20pt
\hsbody=\hsize \hstitle=\hsize 
 
\else\def\apans{h }
\message{(this will be reduced.}
\let\lr=l
\magnification=1000\baselineskip=12pt\voffset=-.31truein\hoffset=-.59truein
\hstitle=7truein\hsbody=3.4truein\vsize=9.5truein\fullhsize=7truein
\ifx\apansw\apans\special{ps: landscape}\hoffset=-.54truein
  \else\voffset=-.25truein\hoffset=-.45truein\fi
\font\titlefnt=amr10 scaled\magstep4 \font\absfnt=amr10 scaled\magstep1
\font\titlemfnt=ammi10 scaled \magstep4\font\absmfnt=ammi10 scaled\magstep1
\font\titlesfnt=amsy10 scaled \magstep4\font\abssfnt=amsy10 scaled\magstep1
 
\output={\ifnum\count0=1 
  \shipout\vbox{\hbox to \fullhsize{\hfill\pagebody\hfill}}\advancepageno
  \else
  \almostshipout{\leftline{\vbox{\pagebody\makefootline}}}\advancepageno
  \fi}
\def\almostshipout#1{\if l\lr \count1=1
      \global\setbox\leftpage=#1 \global\let\lr=r
   \else \count1=2
      \shipout\vbox{\hbox to\fullhsize{\box\leftpage\hfil#1}}
      \global\let\lr=l\fi}
\fi

%

\def\draftmode{\message{ DRAFTMODE }\def\draftdate{{\rm preliminary draft:
\number\month/\number\day/\number\yearltd\ \ \hourmin}}%
\headline={\hfil\draftdate}\writelabels\baselineskip=20pt plus 2pt minus 2pt
 {\count255=\time\divide\count255 by 60 \xdef\hourmin{\number\count255}
  \multiply\count255 by-60\advance\count255 by\time
  \xdef\hourmin{\hourmin:\ifnum\count255<10 0\fi\the\count255}}}

\def\title#1#2{\nopagenumbers\absfnt\hsize=\hstitle\rightline{}%
\centerline{\titlefnt\textfont0=\titlefnt%
\textfont1=\titlemfnt\textfont2=\titlesfnt #1}%
\centerline{\titlefnt\textfont0=\titlefnt%
\textfont1=\titlemfnt\textfont2=\titlesfnt #2
}%
\textfont0=\absfnt\textfont1=\absmfnt\textfont2=\abssfnt\vskip .5in}
 
\def\date#1{\vfill\leftline{#1}%
\tenrm\textfont0=\tenrm\textfont1=\teni\textfont2=\tensy%
\supereject\global\hsize=\hsbody%
\footline={\hss\tenrm\folio\hss}}
%

\def\nolabels{\def\eqnlabel##1{}\def\eqlabel##1{}\def\reflabel##1{}}
\def\writelabels{\def\eqnlabel##1{\hfill\rlap{\hskip.09in\string##1}}%
\def\eqlabel##1{\rlap{\hskip.09in\string##1}}%
\def\reflabel##1{\noexpand\llap{\string\string\string##1\hskip.31in}}}
\nolabels
%
\global\newcount\secno \global\secno=0
\global\newcount\meqno \global\meqno=1
 
\def\newsec#1{\global\advance\secno by1\xdef\secsym{\the\secno.}
\global\meqno=1
\bigbreak\bigskip
\noindent{\bf\the\secno. #1}\par\nobreak\medskip\nobreak}
\xdef\secsym{}
 
\def\appendix#1#2{\global\meqno=1\xdef\secsym{#1.}\bigbreak\bigskip
\noindent{\bf Appendix #1. #2}\par\nobreak\medskip\nobreak}
 
 
\def\eqnn#1{\xdef #1{(\secsym\the\meqno)}%
\global\advance\meqno by1\eqnlabel#1}
\def\eqna#1{\xdef #1##1{(\secsym\the\meqno##1)}%
\global\advance\meqno by1\eqnlabel{#1$\{\}$}}
\def\eqn#1#2{\xdef #1{(\secsym\the\meqno)}\global\advance\meqno by1%
$$#2\eqno#1\eqlabel#1$$}
 
\global\newcount\ftno \global\ftno=1
\def\refsymbol{\ifcase\ftno
\or\dagger\or\ddagger\or\P\or\S\or\#\or @\or\ast\or\$\or\flat\or\natural
\or\sharp\or\forall
\or\oplus\or\ominus\or\otimes\or\oslash\or\amalg\or\diamond\or\triangle
\or a\or b \or c\or d\or e\or f\or g\or h\or i\or i\or j\or k\or l
\or m\or n\or p\or q\or s\or t\or u\or v\or w\or x \or y\or z\fi}
\def\foot#1{{\baselineskip=14pt\footnote{$^{\refsymbol}$}{#1}}\ %
\global\advance\ftno by1}

 
\global\newcount\refno \global\refno=1
\newwrite\rfile
\def\ref#1#2{${[\the\refno]}$\nref#1{#2}}
\def\nref#1#2{\xdef#1{${[\the\refno]}$}%
\ifnum\refno=1\immediate\openout\rfile=refs.tmp\fi%
\immediate\write\rfile{\noexpand\item{\the\refno.\ }\reflabel{#1}#2.}%
\global\advance\refno by1}
\def\addref#1{\immediate\write\rfile{\noexpand\item{}#1}}
 
\def\semi{;\hfil\noexpand\break}




\hyphenation{anom-aly anom-alies coun-ter-term coun-ter-terms}
 


 
   
  \def\bi{{\bf i}}
   
\def\br{{\bf  r}}

\def\pmb#1{\setbox0=\hbox{#1}%
 \kern-.025em\copy0\kern-\wd0
 \kern .05em\copy0\kern-\wd0
 \kern-.025em\raise.0433em\box0 }



\def\cp #1 #2 #3 {{\sl Chem.\ Phys.} {\bf #1}, #2 (#3)}
\def\jetp #1 #2 #3 {{\sl Sov.\ Phys.\ JETP} {\bf #1}, #2 (#3)}
\def\jfm #1 #2 #3 {{\sl J. Fluid\ Mech.} {\bf #1}, #2 (#3)}
\def\jpa #1 #2 #3 {{\sl J. Phys.\ A} {\bf #1}, #2 (#3)}
\def\jcp #1 #2 #3 {{\sl J.\ Chem.\ Phys.} {\bf #1}, #2 (#3)}
\def\jpc #1 #2 #3 {{\sl J.\ Phys.\ Chem.} {\bf #1}, #2 (#3)}
\def\jsp #1 #2 #3 {{\sl J.\ Stat.\ Phys.} {\bf #1}, #2 (#3)}
\def\jdep #1 #2 #3 {{\sl J.\ de Physique I} {\bf #1}, #2 (#3)}
\def\macromol #1 #2 #3 {{\sl Macromolecules} {\bf #1}, #2 (#3)}
\def\pra #1 #2 #3 {{\sl Phys.\ Rev.\ A} {\bf #1}, #2 (#3)}
\def\prb #1 #2 #3 {{\sl Phys.\ Rev.\ B} {\bf #1}, #2 (#3)}
\def\pre #1 #2 #3 {{\sl Phys.\ Rev.\ E} {\bf #1}, #2 (#3)}
\def\prl #1 #2 #3 {{\sl Phys.\ Rev.\ Lett.} {\bf #1}, #2 (#3)}
\def\prsl #1 #2 #3 {{\sl Proc.\ Roy.\ Soc.\ London Ser. A} {\bf #1}, #2 (#3)}
\def\rmp #1 #2 #3 {{\sl Rev.\ Mod.\ Phys.} {\bf #1}, #2 (#3)}
\def\zpc #1 #2 #3 {{\sl Z. Phys.\ Chem.} {\bf #1}, #2 (#3)}
\def\zw #1 #2 #3 {{\sl Z. Wahrsch.\ verw.\ Gebiete} {\bf #1}, #2 (#3)}

\def\br{{\bf r}}

\def\ni{{\noindent}}

\def\pmb#1{\setbox0=\hbox{#1}%
 \kern-.025em\copy0\kern-\wd0
 \kern .05em\copy0\kern-\wd0
 \kern-.025em\raise.0433em\box0 }
\overfullrule=0pt
\font\titlefont=cmbx10 scaled\magstep1

\centerline{\titlefont{An Einstein Model of Brittle Crack Propagation}}

\bigskip
\bigskip

\centerline{Brad Lee Holian$^a$, Raphael Blumenfeld$^{a,b}$, and 
Peter Gumbsch$^c$}

\smallskip
\item{a} Theoretical Division, Los Alamos National 
Laboratory, Los Alamos, NM 87545, USA

\item{b} Cambridge Hydrodynamics, Princeton, NJ 08542, USA

\item{c} Max-Planck-Institut f$\ddot{{\rm u}}$r Metallforschung, 70174 
Stuttgart, GERMANY

\bigskip

We propose a minimal nonlinear model of brittle crack propagation 
by considering only the motion of the crack-tip atom.  The model 
captures many essential features of steady-state crack velocity 
and is in excellent quantitative agreement with many-body dynamical 
simulations.  The model exhibits lattice-trapping. For loads just
above this, the crack velocity  
rises sharply, reaching a limiting value well below that predicted 
by elastic continuum theory.  We trace the origin of the low limiting 
velocity to the anharmonicity of the potential well experienced 
by the crack-tip atom.

\smallskip


\date{}

Recent molecular-dynamics (MD) simulations of crack propagation 
 \ref\hr{B.L. Holian and R. Ravelo, Phys. Rev. {\bf B 51}, 11275 
(1995); S.J. Zhou, P.S. Lomdahl, R. Thomson, and B.L. Holian, 
Phys. Rev. Lett. {\bf 76}, 2318 (1996)}
 \ref\gi{P. Gumbsch, in {\it MRS Symposium Proceedings: 
Fracture -- Instability Dynamics, Scaling, and Ductile/Brittle 
Behavior}, Vol. {\bf 409}, eds. R.L. Blumberg Selinger, J.J. Mecholsky, 
A.E. Carlsson, and E.R. Fuller, jr. (Materials Research 
Society, Pittsburgh, 1996), p.297; see also P. Gumbsch, 
in {\it Computer Simulation in Materials Science}, H.O. Kirchner, 
L Kubin, and V. Pontikis (Kluwer Academic, Dordrecht, Netherlands, 
1996), p.227},
as well as experimental studies 
 \ref\krc{W.G. Knauss and K. Ravi-Chandar, Int. J. Fract. {\bf 27}, 127 
(1985)}
 \ref\fgms{J. Fineberg, S.P. Gross, M. Marder, and H.L. Swinney, 
Phys. Rev. Lett. {\bf 67}, 457 (1991); E. Sharon, S.P. Gross, and 
J. Fineberg, Phys. Rev. Lett. {\bf 74}, 5096 (1993)},
have reflected growing interest in the dynamical aspects of brittle
fracture, including 
the approach to a steady (or quasi-steady) state, the buildup 
of coherent excitation near the crack tip \hr\gi, and the subsequent 
onset of instabilities 
 \ref\mg{M. Marder and S. Gross, J. Mech. Phys. Solids 
{\bf 43}, 1 (1995); E.S.C. Ching, J.S. Langer, and H. Nakanishi, Phys. 
Rev. Lett. {\bf76}, 1087 (1996); E. Sharon, S. P. Gross, and J. Fineberg,
Phys. Rev. Lett. {\bf 74}, 5096 (1995) and {\bf 76}, 2117 (1996)}
 \ref\bi{R. Blumenfeld, Phys. Rev. Lett. {\bf 76}, 3703 (1996); F. Lund,
Phys. Rev. Lett. {\bf 76}, 2724 (1996)}.
In all of these works, it is fair 
to say that a coherent, quantitative understanding of the limiting 
velocity dependence on the local field has not yet been advanced, 
though many good suggestions have been made 
 \ref\sl{K. Sieradzki, G.J. Dienes, A. Paskin, and B. Massoumzadeh, Acta
metall. {\bf 36}, 651 (1988); K. Sieradzki and R. Li, Phys. Rev. Lett.
{\bf 67}, 3042 (1991)}
\gi.  Here, we propose a minimal, one-atom, nonlinear model for
describing brittle fracture, which we call the ``Einstein Ice-Skater''
(EIS) model.

By closely observing movies of MD simulations of brittle 
crack propagation in a two-dimensional (2D) triangular lattice, 
under tensile (transverse, or Mode I) loading and at zero initial 
temperature, we noticed that cracks appear to advance as a sequence 
of essentially one-particle moves.  Along the natural cleavage 
direction separating a pair of close-packed planes (lines in 2D), 
bond-breaking events are well separated in time
 \ref\gzh{P. Gumbsch, S.J. Zhou, and B.L. Holian, Phys. Rev. {\bf B}
(1996, to be published)},
which can be characterized as a zigzag, ice-skating kind of motion between 
the two lines of atoms.  When a bond breaks, the forward partner 
moves ahead, approximately along the former bond direction, while 
the rearward partner swings back to its final equilibrium position 
(see Figure 1).
This led us to speculate that the steady state velocity of 
a brittle crack could be well approximated by a single-particle 
Einstein cell model, where the mobile crack-tip atom (the EIS 
in Figure 1) moves in a field of six immobile neighbors (the sixth, 
with whom the bond has just 
been broken, is assumed to be beyond the range of interaction).  The 
bond-breaking event launches the EIS approximately 
along the bonding direction.  This compressive, nonlinear event 
results in a shearing motion along the transverse pair of close-packed 
lines at $\pm 60^\circ$ to the propagation 
direction, and gives rise to the
local vibrational excitations that build up around 
the crack tip and move coherently with it \hr\gzh.

For sufficiently large strains, the EIS reaches a point that stretches the 
next bond to breaking after a time $t_{\rm break}$ since the last 
bond-breaking event.  The pattern then repeats -- to the other side of the 
ice-skating phase -- and the crack has then advanced by one nearest 
neighbor spacing $r_0$ along the forward direction in the time
$2t_{break}$.  The crack velocity is thus given by
\eqn\Ai{v_{\rm crack} = r_0 / (2t_{\rm break}) \ .} 
To find $t_{\rm break}$, we start from the 
configuration of the EIS and its five connected 
nearest neighbors and solve for the time dependence of the distance 
$r_{01}$ between the EIS and its neighbor 
\#1; $t_{\rm break}$ is the first time that $r_{01}$
reaches the breaking point $r_{\rm max}$.  The equation of motion for the 
position $\br_0$ of the EIS (atomic mass $m$) is
\eqn\Aii{m\ddot{\br_0} = 
- \sum_{i=1}^6 \partial\phi(r_{0i})/\partial\br_0\ ,}
which can be solved given the pair potential 
$\phi(r_{0i})=\phi(|\br_0-\br_i|)$ and the initial conditions.  This equation 
is not trivial to solve, even for harmonic potentials, but can be solved 
numerically.  We first assume the initial EIS coordinates $x = a/2$, $y = 0$ 
and velocities $\dot x = \dot y = 0$ (the initial velocities of
steady-state crack-tip atoms in full MD simulations are observed to be
indeed very small). With $r_o \equiv 1$, the six immobile neighbors are assumed 
to be located at 
($- a/2, 1/2$), ($a/2, 1$), ($3a/2, 1/2$), ($3a/2, - 1/2$), 
($a/2+a_0\epsilon, -1/2$), and ($- a/2 - a_0\epsilon, -1/2$), where 
$a_0 = \sqrt{3}/2$, $a = a_0(1+\epsilon)$, and $\epsilon$ is the uniaxial 
strain in the transverse direction to crack propagation.  (See Figure 1.)

We can obtain a crude estimate for $t_{\rm break}$ by imagining that the EIS 
starts at the turning point of its motion in the final harmonic equilibrium 
well.  The bulk Einstein model is characterized by a frequency 
of $\omega_E = \sqrt{3}\omega_0$, where $\omega_0$ is the fundamental 
frequency given by $m\omega_0^2 = \phi''(\br_0)$.  Hence, 
if the time $t_{\rm break}$ is one-half the 
period (from one turning point to the other at bond breaking), 
then $v_{\rm crack} =\sqrt{3}r_0\omega_0/2\pi$.  Since the 
triangular-lattice shear-wave speed $c_{\rm s} = \sqrt{3/8}r_0\omega_0$ 
(which is very close to the Rayleigh, or surface wave speed
 \ref\ah{W.T. Ashurst and W.G. Hoover, Phys. Rev. {\bf B 14}, 1465 (1976)}), 
$v_{\rm crack}/c_{\rm s}\approx\sqrt{2}/\pi  = 0.45$, independent of the 
anharmonicity of the potential.  Since the effective frequency 
of a stretched anharmonic bond decreases (actually to zero at 
the inflection point), the crack velocity in the anharmonic case 
should be lower.

To go beyond this estimate, we investigated two kinds of attractive 
snapping-bond potentials: harmonic (HSB) and anharmonic (ASB), the latter 
based on the Morse potential
\eqn\Aiii{\phi(r) = \left[1 - e^{-\alpha(r-1)}\right]^2/(2\alpha^2)\ .} 
Here we scale the distance by $r_0$ and the energy by $mr_0^2\omega_0^2$; 
$\alpha$ is the repulsive parameter (the familiar Lennard-Jones 6-12 
potential is closely approximated 
by $\alpha  = 6$; most materials can be represented by $4 \le \alpha \le 6$). 
The ASB potential is obtained from Eq. $\Aiii$ 
by flattening it out at $r_{\rm max} = 1 - \ln{(1 -\sqrt{\chi})}/\alpha$ in 
the attractive region (beyond the minimum of the Morse potential); the 
cohesive energy is then $\chi/(2\alpha^2)$, 
for $\chi < 1$.  The ASB force jumps discontinuously at this 
point from a negative value to zero -- hence the term ``snapping-bond.'' 
For small displacements about $r = 1$, Eq. $\Aiii$ is approximately 
harmonic, $(r - 1)^2/2$.  The HSB potential cuts off at the same energy as 
the ASB, but at $r^0_{\rm max} = 1 + \sqrt{\chi}/\alpha < r_{\rm max}$.
We find that {\it the range and maximum attractive force of the 
potential are the essential parameters that govern the crack velocity}.

Our choice of snapping-bond potentials makes precise the definition of the 
distance beyond which a bond is considered ``broken,'' an ambiguous concept 
for completely continuous potentials.  Since our goal is to compare 
this EIS model with a fully dynamical system, a well-defined breaking 
point for both is a distinct advantage.  The fully dynamical systems we 
compare with are rather restrictive, namely, close-packed 
lines of atoms of width $w=4,8,16$, and $64$, with the outer two clamped, 
and the inner free to move; moreover, only nearest-neighbor interactions are 
considered. (Strips were typically $200r_o$ in length; steady-state
propagation is attained well within 10\% of that length.)

For this thin-strip, fixed-grip geometry, the critical Griffith strain 
$\epsilon_G$ for initiating forward crack motion can be computed by equating 
the potential energy in two transverse sections of the strip of height 
$r_0/2$: one far behind the crack with all 
bonds in equilibrium, except for the one broken bond, and the 
other far in front, with all bonds equally stretched.  The Griffith 
criterion is obtained from
\eqn\Aiii{(w-1)\phi(r_1)=\phi(r_2)=\chi/(2\alpha^2)\ ,} 
where $r_1$ is the elastically stretched 
bond ($r_1^2=a^2+1/4$) and $r_2$ is the broken bond across the gap of the 
relaxed crack.  The Griffith criterion $\epsilon_G$ is thus
\eqn\Av{
\epsilon_G = \left\{
\matrix
{ \left\{1 - {8\over{3\alpha}}\ln{\left(1 - \sqrt{{\chi\over{w-1}}}\right)}
\left[1 - {1\over{2\alpha}}\ln{\left(1 - \sqrt{{\chi\over{w-1}}}\right)}\right]
\right\}^{1/2} - 1 & , & {\rm ASB} \cr
\left\{1 + {8\over{3\alpha}}\sqrt{{\chi\over{w-1}}}
\left[1 + {1\over{2\alpha}}\sqrt{{\chi\over{w-1}}}\right]\right\}^{1/2} - 1 
\hskip1.12in & 
, & {\rm HSB} \ .\cr} \right.
}

An intriguing aspect of the EIS model is the straightforward 
emergence of the lattice trapping phenomenon
 \ref\thom{R. Thomson, Solid State Phys. {\bf 39}, 1 (1986)}:
unless the strain exceeds a value well above $\epsilon_G$, the distance 
between the EIS and its neighbor \#1 will not reach $r_{\rm max}$.  The 
strain must therefore exceed $\epsilon_G$ by a barrier amount of overstrain 
that is a characteristic of the atomistic nature of the crack 
tip, and which can only be evaluated atomistically. In Figure 2, 
we show our results for the crack-tip 
velocity (in units of $c_{\rm s}$), as a function of the 
strain, for the EIS 
model and for 
the fully dynamical $w=4$ strip ($\alpha = 6$, $\chi = 1/2$).  The EIS model 
agrees to within 10\% of the velocity with the MD results --
remarkable for such a simple model.

However, the lattice-trapping strain is underestimated by 13\% for
the anharmonic and 12\% for the harmonic system, which is most clearly due to 
neglected correlations with farther neighbors in the EIS model. 
For the anharmonic system, the onset of crack motion for the 
fully dynamical $w=4$ strip occurs at a crack velocity of about 
30\% of the shear-wave speed, while for the harmonic system, the 
crack starts at about 50\% of the shear-wave speed.  Under further 
loading, the crack-tip velocity increases roughly linearly with 
strain but with a higher slope for the harmonic than for 
the anharmonic system.

To compare our EIS results to MD simulations and experiments  
we rescaled the wider system strains by the Griffith strain 
($\epsilon_G\sim 1/\sqrt{w}$) and found 
good agreement, except for slight, but systematic 
increases in the lattice trapping strain with size for harmonic potentials. 
We can understand this by noting that wide 
anharmonic systems, where stretched bonds weaken, are more compliant and 
tend to have local strains near the crack tip that are closer to those in
the narrow strips. On the other hand, harmonic bonds do not weaken with
stretching, so that the global strain is spread more democratically
across the system.  We emphasize that, even in wide systems where 
the global strain can be arbitrarily small, the fact that local 
strains near the crack tip are large (of order 10\%, as in the 
narrow-strip case) is a significant reason for the success of the 
EIS model.

We find that crack velocities in anharmonic systems are essentially 
independent of the 
anharmonicity parameter, at least 
over the range $4\le\alpha\le 6$; in fact, the curves 
for $\alpha=4$ and 5 practically overlap.  As the cohesive 
strength $\chi$ decreases from 1/2 down to 1/8 (along with the range of
the potential), crack velocities in anharmonic systems show a slight
increase ($\sim10\%$) in ultimate slope 
and greater variability in the jump-off lattice-trapping strain. (In the
limit $\chi \rightarrow 0$, of course, the harmonic limit is approached
[7].) In general, velocities in anharmonic systems are lower than in 
harmonic ones, show less variation 
with strain, and exhibit relatively lower lattice trapping (when the 
strain is scaled by $\epsilon_G$).  Similar trends are 
exhibited in the full MD simulations, including those using full, 
continuous (rather than discontinuous snapping-bond) 
potentials \hr\gi\sl\gzh, and those for 
systems much wider 
than $w=4$. Again, the principal differences are in the lattice-trapping 
strains.  We can therefore conclude that the EIS approximates 
very well the crack-tip atomic motion, just as our intuition from 
larger-scale MD simulations had suggested.

Our minimal EIS model indeed confirms speculations about 
the correlation of the limiting steady-state crack-tip velocity 
and anharmonicity \gi\sl\gzh.  The more ``realistic'' 
anharmonic interactions give steady-state crack-tip 
velocities that never exceed $0.4$ of the Rayleigh speed, in excellent 
agreement with experimental observations \krc\fgms.  With the EIS 
model, the origin of this low speed can clearly be attributed 
to the smaller attractive force on the crack-tip atom at the point 
of bond breaking, as compared to the harmonic (or linear elastic) 
analysis.

Under loading, the thin-strip MD crack-tip velocity in Figure 
2 jumps sharply at the lattice-trapping strain to a slowly rising 
linear regime, and then once again rises sharply at a strain of 
$0.15=1.3\epsilon_G$.  Close examination of atomic 
configurations revealed that this second rise is associated with 
two instabilities: the first is a wake of large-amplitude surface 
(Rayleigh) waves behind the crack tip; at somewhat higher strains, 
the crack begins to jump from the central channel to one of the 
side channels next to the fixed-grip atoms (see Figure 1).  We 
have observed this zigzag 
propagation by plus or minus one channel in much wider systems, 
where, at even higher strains, dislocations are emitted, followed 
immediately by branching.  Dynamical instabilities such as these 
divert energy from brittle bond breaking, causing the crack-tip 
velocity to drop rather than rise.  Dislocation emission and real 
crack branching are, of course, forbidden processes in the artificially 
narrow 4-wide strip, and are completely absent in the one-particle 
EIS model.

Finally, the hysteresis under unloading and healing up of 
the crack can be obtained from the EIS.  To do this, we simply 
detect when the 6-neighbor model reconnects the bond between the 
EIS atom and neighbor \#6, rather than opening up the crack in 
the forward direction.  This occurs soon below $\epsilon_G$ for the 
anharmonic potential (namely, $0.98\epsilon_G$), but substantially lower 
for the harmonic potential ($0.85\epsilon_G$).  Crack propagation 
and crack healing are thus quite asymmetric processes.

In conclusion, the Einstein Ice-Skater model of brittle crack propagation 
is able to predict quantitatively the steady-state crack velocity 
under loading, including lattice trapping, as well as hysteresis 
upon unloading and crack healing.  The maximum velocity achieved 
in full MD simulations as a function of strain is principally 
limited by the anharmonicity in the attractive region of the pair 
potential, which is captured by the EIS; however, it is also affected 
by instabilities that involve collective motion (energy buildup, 
dislocation emission, 
and branching), which is inaccessible to the one-particle EIS model. 
Nevertheless, this simple EIS model allows us to
explain, in quite satisfactory quantitative fashion, the effect of
nonlinear motion of the crack-tip atom on the limiting crack velocity.

\smallskip

\ni We thank Robb Thomson, Shujia Zhou, and 
Bill Hoover for stimulating discussions. 

\bigskip

\centerline{\bf FIGURE CAPTIONS}

\ni \item{1.} Initial atomic coordinates for crack propagation in a 
triangular-lattice strip, four close-packed lines wide; the outer two 
lines of atoms are fixed, while the inner two are mobile.  Heavy lines 
indicate equilibrium (nearest-neighbor) bonds of length $r_0 = 1$; 
heavy dashed lines are slightly stretched, nearly vertical 
bonds; light lines are bonds elastically stretched to length 
$r_1\approx 1 + 3\epsilon / 4$ by the uniaxial 
strain $\epsilon$ in the $x$-direction; the light dashed line is 
a just-broken bond with neighbor \#6.  The EIS atom is indicated 
by the large open circle: it moves initially approximately in the direction 
of the arrow, stretching the bond with neighbor \#1 until breakage, 
then heads toward its final equilibrium position (small circle).

\ni \item{2.} Crack velocity (in units of shear-wave speed $c_s$) as a 
function of strain for the anharmonic snapping-bond (ASB) and 
harmonic snapping-bond (HSB) potentials.  Results for the EIS model are 
shown for Morse parameter $\alpha=6$ and cohesive bond-strength 
$\chi=1/2$, along with $w=4$ strip MD simulations (closed circles 
for ASB and open for HSB).

\vfill\immediate\closeout\rfile
\baselineskip=18pt\centerline{{\bf REFERENCES}}\bigskip\frenchspacing%
\input refs.tmp\vfill\eject\nonfrenchspacing

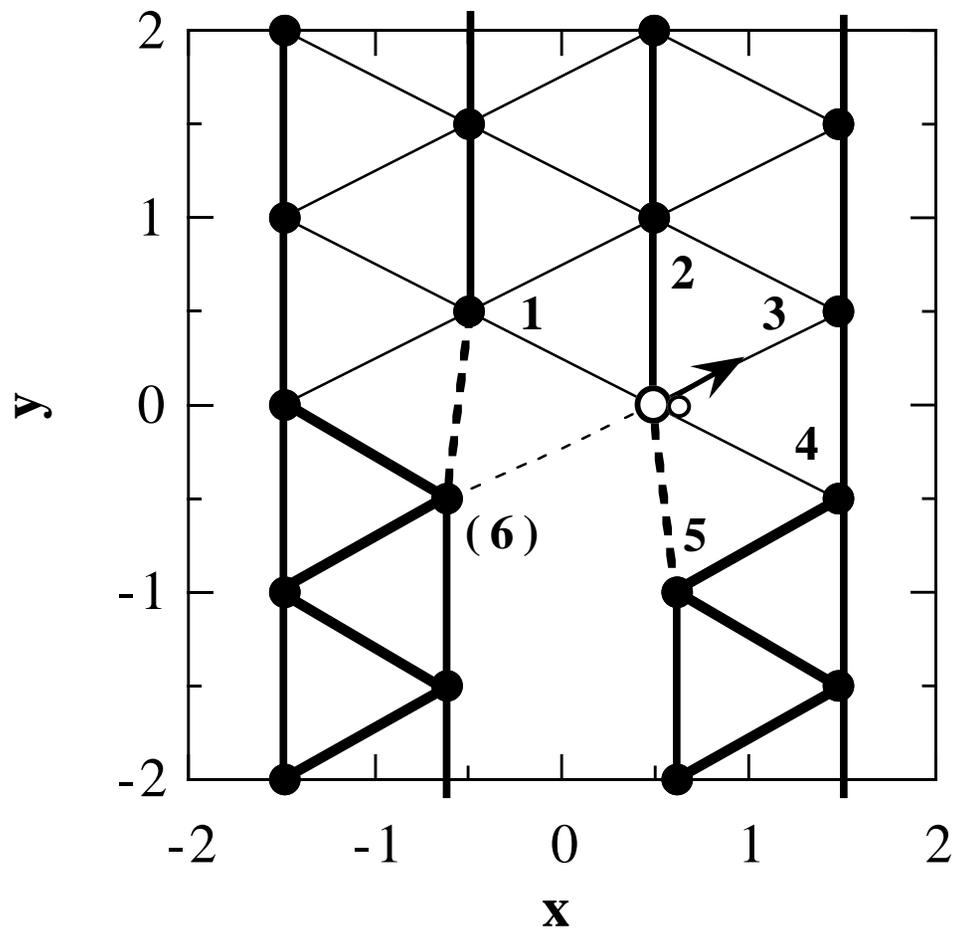
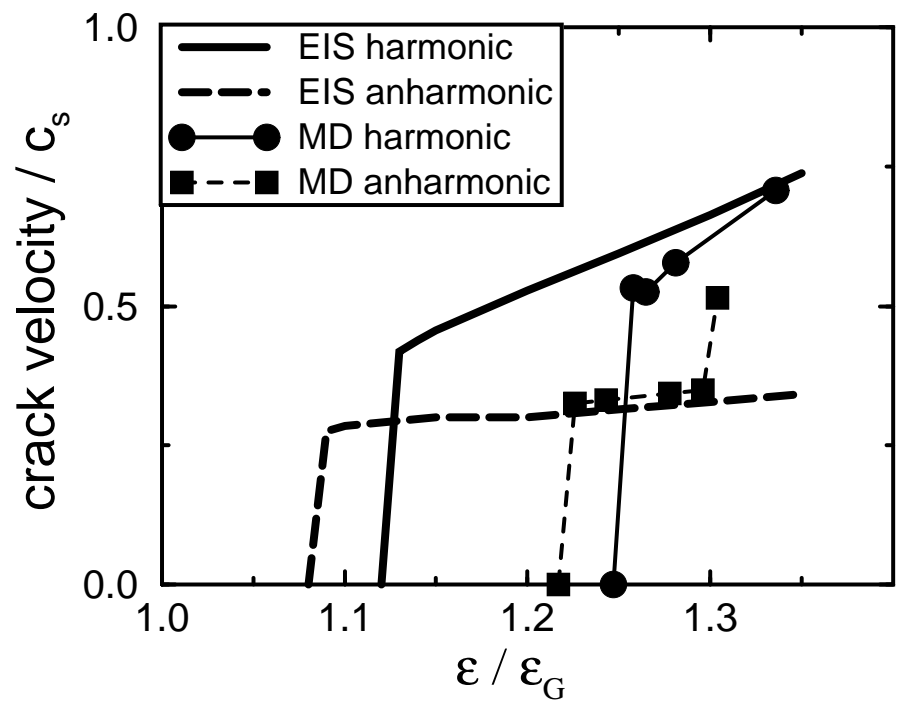
\end